\documentclass[aps,twocolumn,amsfonts,showpacs]{revtex4}
\usepackage{epsfig,amsopn,amsmath}
\usepackage{multirow,bigdelim}

\usepackage{color}
\usepackage{lipsum}

\begin{document}

\title{Coexistence of competing metabolic pathways in well-mixed populations}

\author{Lenin Fernandez}
\affiliation{%
Departamento de F\'{\i}sica, Universidade Federal de Pernambuco, \\
52171-900 Recife-PE, Brazil
}%
\author{Andr\'e Amado}
\affiliation{%
Departamento de F\'{\i}sica, Universidade Federal de Pernambuco, \\
52171-900 Recife-PE, Brazil
}%
\author{Fernando Fagundes Ferreira}
\affiliation{%
Escola de Artes, Ci\^encias e Humanidades, Universidade de S\~ao Paulo,
03828-000 S\~ao Paulo, Brazil
}%
\author{Paulo R. A. Campos}\email{prac@df.ufpe.br}
\affiliation{%
Departamento de F\'{\i}sica, Universidade Federal de Pernambuco, \\
52171-900 Recife-PE, Brazil
}%

\begin{abstract}
Understanding why strains with different metabolic pathways that compete for a
single limiting resource coexist is a challenging issue within 
a theoretical perspective.
Previous investigations rely on mechanisms such as group or spatial
structuring to achieve a stable coexistence between competing metabolic
strategies.
Nevertheless, coexistence has been experimentally reported even 
in situations where
it cannot be attributed to spatial effects [Heredity {\bf 100}, 471 (2008)].
According to that study a toxin expelled by one of the strains can be
responsible for the stable maintenance of the two strain types.
We propose a resource-based model in which an efficient strain 
with a slow metabolic rate competes with a second strain type which 
presents a fast but inefficient metabolism. Moreover, the model assumes that the inefficient strain produces a toxin as a byproduct.
This toxin affects the growth rate of both strains with different strength.
Through an extensive exploration of the parameter space we determine the situations at which the coexistence of the
 two strains is possible. Interestingly, we observe that the resource influx rate plays a key role in the maintenance 
of the two strain types. In a scenario of resource scarcity the inefficient is favored, though as the resource influx rate 
is augmented the coexistence becomes possible and its domain is enlarged.
\end{abstract}


\maketitle

\section{Introduction}\label{section}
According to the evolutionary theory the main mechanisms driving the evolution of natural populations are Darwinian selection, 
genetic drift, mutation and migration \cite{fisher1930genetical, Haldane1927, Wright1931}. However, these mechanisms alone do not explain the emergence of more complex life forms from simpler units.
The increase in complexity at the organism level is supposed to be related to the emergence of cooperation \cite{Hamilton1964}. 
Cooperation goes against the nature of the individuals who are supposed to  act selfishly and favor their own genes.

The maintenance of the cooperative behavior is still an intriguing and 
open topic in evolutionary biology \cite{AxelrodBook1984}. One individual 
is said to display cooperative behavior if 
it provides a benefit to another individual or to a group at the expense of 
its own relative fitness. On the other hand, in a defecting behavior the recipient gets the benefit of the interaction without reciprocity and paying no cost for the action. 
The interaction of RNA phage $\phi 6$, a viral genotype 
that synthesises large quantities of products as a common good, and its mutant $\phi H2$, a genotype that synthesises 
less but specialises in sequestering a larger share of the products made by the others, can be described as a cooperator-defector relationship \cite{TurnerNature1999}.
Another example is seen in the evolution of metabolic pathways \cite{PfeifferScience2001}. Pfeiffer et al. observed similar relationship when studying the trade-off between yield and rate in heterotrophic organisms \cite{PfeifferScience2001,PfeifferTrends2005}.

The main source of energy for survival and reproduction of heterotrophic organisms comes from ATP (adenosine triphosphate). The basic raw material for the synthesis of ATP is glucose. The conversion of glucose in ATP occurs mainly through one of two metabolic pathways: fermentation and respiration \cite{RichBST2003,HellingJBac2002}.
Although respiration is more efficient, i.e., more energy produced per glucose unit (high yield), it is slower than fermentation. 
In its turn, fermentation inefficiently produces ATP but can achieve high 
rate of growth (high rate) at the expense of depletion of the resource. 
Therefore the conversion of resource into ATP, and consequently growth, is driven by a trade-off between yield and rate \cite{MacleanHeredity2008}.

This trade-off gives rise to a social conflict \cite{HardinScience1968}. 
Empirical studies demonstrate that the trade-off between resource uptake and yield is a common place in the
microbial world \cite{MeyerNatComm2015,KapplerMicrobiology1997,PfeifferScience2001,OtterstedtEMBO2004}, which occurs due 
to biophysical limitations preventing organisms to optimize 
multiple traits simultaneously. Independent experiments have reported the existence of a negative correlation between rate and
yield \cite{MeyerNatComm2015,NovakAmNat2006,LipsonBiogeo2009,PostmaApEnvMic1989}.
The respiration mode of processing glucose was made possible about 2.5 billion years ago during the Great Oxygenation Event that introduced free oxygen in the atmosphere \cite{DonoghueNature2010}. One important issue is to understand under which conditions the efficient mode of metabolism, which displays a lower growth rate, could arise and fixate.
Previous studies have tried to understand the mechanism that can favor the 
fixation and maintenance of the efficient strain. In spatially homogeneous 
populations the most influential factor determining the fate of the 
population is the resource influx rate. Under resource scarcity the 
efficient strain tends to dominate, while under the scenario of abundant 
resource the inefficient strain thrives \cite{Frick2003}. Besides, it has 
been shown that under the scenario of spatially structured populations or 
populations structured in groups a more favorable scenario for the fixation 
of the efficient trait is created \cite{PfeifferScience2001, 
PfeifferTrends2005, AndreCampos2015, Aledo2007}. Following those investigations the coexistence is found 
under specific conditions in spatially 
structured populations \cite{Aledo2007}, but not found in well-mixed populations.
Though, an experimental study with yeast populations in batch culture showed 
that, in the presence of a toxic metabolite, coexistence can be achieved \cite{MacLeanNature2006}.

In the current work, we survey the conditions under which a toxic 
metabolite can promote a stable coexistence of two strains that  differ in 
their metabolic properties. This study is carried out by finding the solutions and performing a stability analysis of a discrete-time model.

\section{The Model}\label{section}
A well-mixed population of variable size with two competing strains is considered. 
The competing strains are described as cooperator, denoted by $C$, or defector, represented by $D$. 
The former makes efficient use of resource, as it  converts resource into ATP at high yield. Though strain $C$ has a low
uptake rate of resource. On the other hand, strain $D$, defector, is characterized by a rapid metabolism, 
and hence it can consume resource at high rate, however its machinery of conversion of resource into
ATP is inefficient. The influx of resource into the system (e.g. glucose), hereafter $S$, is kept constant over time 
and directly influences the population size at stationarity.  The population size is not uniquely determined by
the resource influx rate but also by the population composition, as the strains have distinct metabolic properties, thus  
 changing the rate at which the cells divide.

At each generation every individual goes through the following processes: resource uptake, conversion of the caught resource into 
energy, which can lead to cell division, and stochastic death. 
First there is a competition for the resource. At this stage, strains of type $D$, characterized by a rapid
metabolism, are stronger competitors, as they present a larger rate of consumption and  thus seize 
a larger portion of the shared resource than strains type $C$.  The amount of resource captured by each strain of type $D$ is
\begin{align}\label{J}
S_D(S) = \frac{A^S_D S}{A^S_D N_D+A^S_{C} N_C},
\end{align}
whereas 
\begin{align}\label{J1}
S_C(S) = \frac{A^S_CS}{A^S_D N_D+A^S_{C} N_C}.
\end{align}
denotes the  amount of resource captured by a strain of type $C$. The quantities
$A_{C}$ and $A_{D}$ are, respectively, the consumption rates of strains $C$ and $D$, while $N_C$ and $N_D$ are
their population numbers. From eqs. (\ref{J}) and (\ref{J1}) follows that $S_D N_ D + S_C N_C=S$, as required.
Since by definition a defecting strain displays a larger consumption rate,  $A_D > A_C$. Empirical studies suggests that
the consumption rate of defectors is ten-fold higher than the consumption rate of cooperators \cite{OtterstedtEMBO2004}.

In the subsequent process, the resource is then converted into energy (ATP), and so
increasing the individuals' energy storage. The increase in the internal energy of a given individual $j$, $E_{j}$,  of type
$k=C,D$, is given by  
\begin{align}\label{energy}
\Delta E_{j} = J^{ATP}_k(S_k) 
\end{align}
The functions $J_k^{ATP}(S_k)$ with $k=C,D$ determine how efficiently energy is produced from the
captured resource for each strain type. As functional responses $J^{ATP}_i$, $i=C,D$, we choose a Holling's type II function,  which displays a decelerating rate of conversion of
resource into energy and follows from the assumption that the consumer is restricted by its capability to process the resource. 
As such, we propose as functional responses
\begin{subequations}\label{parametrization} 
\begin{align}
	&J^{ATP}_C(S_i^C) = A^{ATP}_C\left(1-\exp(-\alpha^{ATP}_C{S_C})\right)\\
	&J^{ATP}_D (S_i^D)= A^{ATP}_D\left(1-\exp(-\alpha^{ATP}_D{S_D})\right).
\end{align}
\end{subequations}
As they should be, the functions depend on the amount of seized resource, $S_{i}$, $i=C,D$, and 
the exponents $\alpha^{ATP}_{i}$, $i=C,D$, tune
the efficiency of the process of conversion of resource into energy. The smaller the $\alpha^{ATP}$ is, the more inefficient
the metabolic pathway becomes. 

As aforementioned, the efficient strain $C$ converts resource into internal energy efficiently but at a low consumption rate, 
while the one that metabolizes fast, strain 
 $D$, exhibits an opposed behavior. The above scenario is simulated by holding $A_D > A_C$, assuring that strategy $D$ 
has a larger uptake rate, together with the condition $\Delta_{ATP}=\alpha^{ATP}_D/\alpha^{ATP}_C < 1$. The latter requirement
aims to warrant that the strategy $D$ is less efficient in producing energy from the resource. 
If $\alpha^{ATP}$ is large even a small amount of resource already provides
nearly the maximum amount of energy that can be carried out in one life cycle.  
The fact that the rapid metabolic pathway is inefficient does not ensue that the strain $D$ will not end up producing
a amount of energy higher than strain $C$. Indeed, this situation can be achieved in case
a large amount of resource is captured. As an example we mention unicellular eukaryotes like yeast 
 \cite{Frick2003, OtterstedtEMBO2004}, which concomitantly uses two different pathways 
of ATP production, working as a respiro-fermenting cell. Therefore it is of particular interest the 
condition $\Gamma_{ATP} = A^{ATP}_D/A^{ATP}_C > 1$, which allows the strain $D$ to end up with a larger 
amount of energy at the expense of a big amount of
resource. Under this situation, strain type  $C$ faces the worst environmental situation and in the case
it can be selected for in such conditions it will certainly thrive in more  favourable scenarios.

\subsection{Cell division and death}\label{sub}

The number of individuals (cells) is not fixed but variable over time, being determined by the 
intrinsic dynamics of the population and the influx of resource
into the system.  Every generation,  each cell divides into  two identical cells whenever
its energy storage $E_{i}$ (group index, $i=1,..., N$) reaches a threshold value, $E_{\rm{max}}$. 
The daughter cells are endowed with half of the energy of the parental cell.

The model assumes that individuals can also die spontaneously with a small probability  $\nu$ per generation.  The model also assumes that
the growth rates of both strain types are affected due to byproducts (toxin) produced by the  inefficient strain. The 
effect of the toxins on strains $C$ and $D$ are not the same being simulated as a reduction in the growth rate 
which is density dependent, $\eta N_{D}$ and $\beta N_{D}$ for
strains $D$ and $C$, respectively.

The assumption that only the inefficient strain produces the toxin is quite reasonable since the fermentation process 
gives rise to many byproducts, such as alcohol and acethic acid, which carry substantial amount of free energy. On the other hand,
the respiration process leads to the production of water and carbon dioxide, which are easily released by the cell.

\section{Results}\label{section}
As aforesaid, the model considers two types of strains: strain $C$, which displays a high yield in energy production, and a second 
strain which is characterized by a high uptake rate of resource though achieves a lower yield in the process of energy conversion. 
Additionally, the strain $D$ generates the toxin that directly reduces the growth rate of the two strain types. 

As an individual replicates at a rate which is proportional to the rate the cell generates ATP, in a discrete-time model
formulation the population numbers, $n_D$ and $n_{C}$, can be written as 
\begin{subequations}\label{modeleq}
\begin{align}
	n_D(t+1) &= n_D(t) \left[1+a_D \left(1-\text{e}^{-\alpha_D S_D}\right)-\nu -\eta\, n_D(t)\right]\\
	n_C(t+1) &= n_C(t) \left[1+a_C \left(1-\text{e}^{-\alpha_C S_C}\right)-\nu -\beta\, n_D(t)\right].
\end{align}
\end{subequations}
In the above equations $n_i(t)$, $i=C,D$ represents the population size of strain type $i$ at time $t$, and $a_i=A_{i}^{ATP}/E_{\rm{max}}$ is 
a necessary rescaling as a cell only divides after its energy storage surpasses the energy threshold $E_{\rm{max}}$. The other 
parameters are as defined before. 

The set of equations (\ref{modeleq}) has several solutions. As standard, a solution is the set of values $(\hat{n}_{D},\hat{n}_{C})$
that satisfies the conditions $n_{D}(t+1)=n_{D}(t)=\hat{n}_{D}$ and $n_{C}(t+1)=n_{C}(t)=\hat{n}_{C}$.
The simplest ones are those corresponding to isogenic populations at the steady state, i.e., either strain $D$ or strain $C$ remains with
the exclusion of the other one. The extinction of both strain types can also be achieved under very simple assumptions, as discussed later.
Of particular interest to us, is the solution that guarantees a stable coexistence of both strain types. Actually, this is the greatest
motivation of the current study. Below we will present the different situations and possible solutions.  A stability analysis 
to delimit the region
of the parameter space at which the solutions are stable is also carried out. The possible solutions of eqs. (\ref{modeleq}) are:
I) $\hat{n}_{D}=0$ and $\hat{n}_{C}=0$; II) $\hat{n}_{D}=0$ and $\hat{n}_{C} \neq 0$;  III) $\hat{n}_{D} \neq 0$ and $\hat{n}_{C} = 0$;
and the coexistence solution IV)  $\hat{n}_{D} \neq 0$ and $\hat{n}_{C} \neq 0$.

\begin{figure*}[tp]
\centering
	\includegraphics[width=0.9\textwidth]{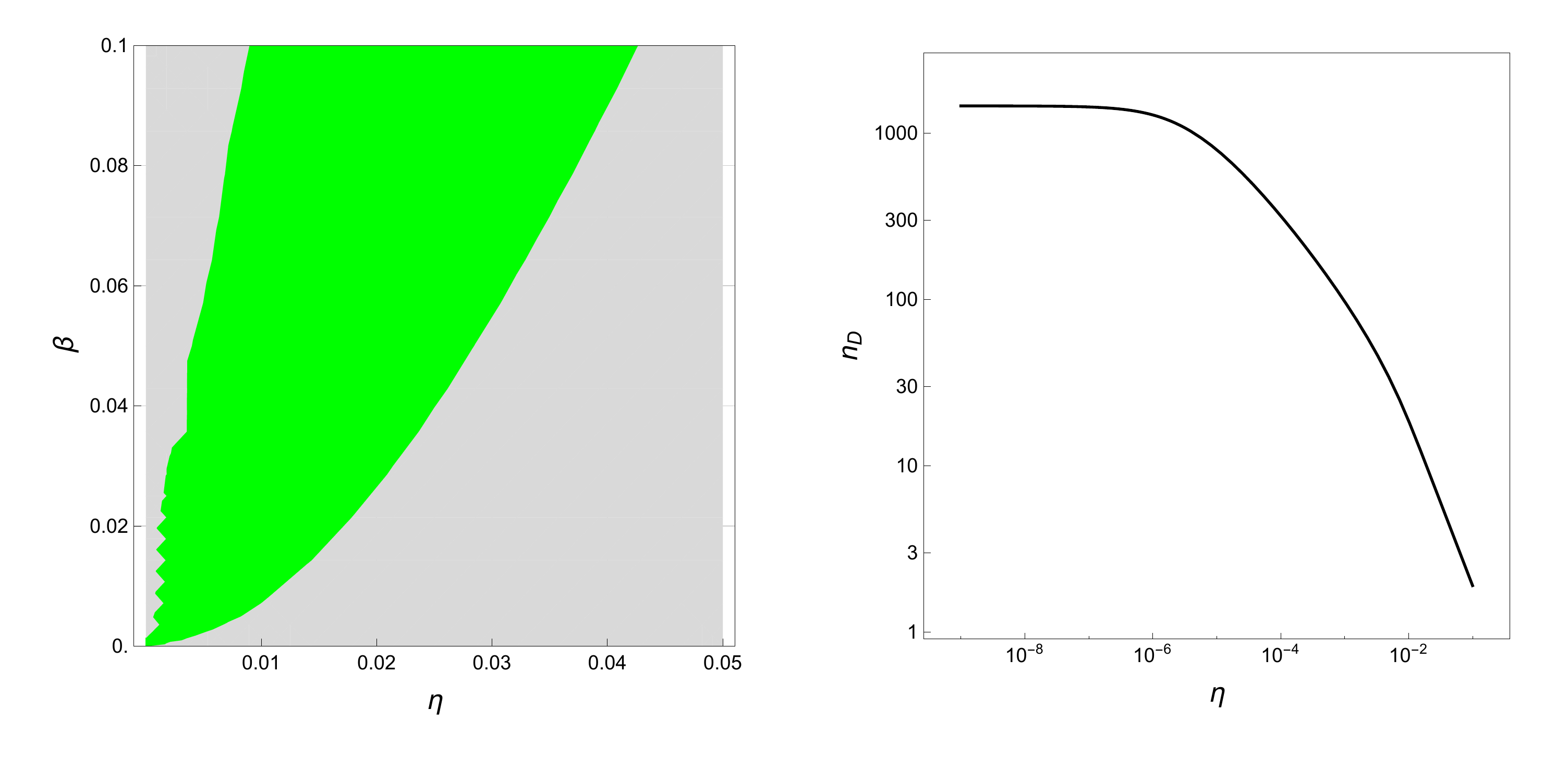} \caption{Left graph: region where the solution with only strain 
type $D$ is stable (in green) as a function of $\eta$ and $\beta$. The grey area denotes the region where the solution exists but it is not stable. Right graph: population size for a situation with strain $D$ only as a function of $\eta$. 
This graphs were obtain for the parameters $\nu = 0.01$, $\Delta = 0.5$, $\Gamma = 4$, $a_D = 0.2$, $S=150$ and $\alpha_D = 0.5$. 
{\bf What is the value of $\beta$ for the right panel? $\beta$ is arbitrary because $n_D$ doesn't depend on $\beta$ when $n_C=0$}}
\label{fig:Donly}
\end{figure*}

\subsection{Solution I: $\hat{n}_D=0$; $\hat{n}_C=0$}\label{subsub}
A trivial solution is $\hat{n}_D=0$; $\hat{n}_C=0$. Calculating the eigenvalues of the Jacobian matrix of the system we verify that this solution is stable only when $\nu>a_D$ and $\nu>a_C$. These conditions just mean that extinction is reached when the death rate is 
larger than the growth rates of both strain types.

\subsection{Solution II: $\hat{n}_D=0$; $\hat{n}_C \neq 0$}\label{subsub}
If the strain $D$ inexists at the stationary regime, i.e.  $\hat{n}_D = 0$, it can be easily shown that 
\begin{align}\label{solution-C}
	\hat{n}_C = -\frac{\alpha_C\, S}{\log \left(1-\frac{\nu }{a_C}\right)}.
\end{align}
The above solution is the same found in Ref. \cite{AndreCampos2015}, as in the absence of defectors no toxin is produced.
An evolutionary invasion analysis enables us to determine the conditions under which the solution is evolutionary stable, i.e., 
once the efficient trait $C$ is established it can not be invaded by  the defecting trait $D$. In order to accomplish this,
the Jacobian of the system (\ref{modeleq}) is evaluated 
at  $\hat{n}_{D}=0$ and $\hat{n}_{C}=-\frac{\alpha_C\, S}{\log \left(1-\frac{\nu }{a_C}\right)}$.  Therefore, the Jacobian matrix
becomes
\begin{widetext}
\begin{align}
J_{n_D(t)=0}=	\left(
\begin{array}{cc}
 1-\nu + a_D \left[1-\left(1-\frac{\nu }{a_C}\right)^{\epsilon \Delta}\right] & 0 \\
\frac{S\, \alpha_C\, \beta }{\log \left(1-\frac{\nu }{a_C}\right)}
+\epsilon a_C(1-\frac{\nu}{a_C} ) \log \left(1-\frac{\nu }{a_C}\right) & 
 1 + \left(a_C-\nu\right) \log \left(1-\frac{\nu }{a_C}\right)
\end{array}
\right)\label{jacobiannd0}
\end{align}
\end{widetext}
The stability of the solution is determined by the eigenvalues of the Jacobian matrix.  Due 
to its form (eq. \ref{jacobiannd0})), the 
eigenvalues 
are simply given by its diagonal elements, i.e.
\begin{align}
	\lambda_1 &= 1-\nu + a_D \left[1-\left(1-\frac{\nu }{a_C}\right)^{\epsilon \Delta}\right] \label{eigenvalue1}\\
	\lambda_2 &= 1 + \left(a_C-\nu\right) \log \left(1-\frac{\nu }{a_C}\right) \label{eigenvalue2}
\end{align}
where $\epsilon \equiv A_D/A_C$ and $\Delta\equiv\alpha_D/\alpha_C$. To be stable, a solution in a discrete-time formulation 
requires that $|\lambda_i| < 1$ for $i=1,2$. If we assume the biologically plausible assumption that $\nu/a_C \ll1$, these 
eigenvalues can be approximated as
\begin{align}
	\lambda_1 &\approx 1-\nu + a_D \left[1-\left(1-\frac{\nu \epsilon \Delta}{a_C}\right)\right] = 1-\nu\left(1-\epsilon \Delta\Gamma\right)\label{eigenvalue1-1}\\
	\lambda_2 &\approx 1 + \left(a_C-\nu\right) \left(-\frac{\nu }{a_C}\right)=1-\nu\left(1-\frac{\nu}{a_C}\right)  \label{eigenvalue2-1}
\end{align}
where $\Gamma\equiv a_D/a_C$.

As $|\lambda_i| < 1$ is needed to assure stability of the solution, looking at the eq. (\ref{eigenvalue1-1}) some restrictions on 
the parameter  values are settled:
\begin{align}
	\lambda_1 < 1  &\Rightarrow \epsilon \Delta\Gamma < 1\label{firstcondition}\\
	\lambda_1 > -1 &\Rightarrow \nu\left(1-\epsilon \Delta\Gamma\right) < 2.
\end{align} 
The second condition is always automatically verified as $\epsilon \Delta\Gamma>0$ and $\nu<1$. On the other hand, the 
condition (\ref{firstcondition}) imposes that $\epsilon \Delta\Gamma < 1$ in order to turn the population of efficient strains
evolutionarily stable against invasion of the selfish strain $D$. 
The condition matches the one derived in Ref. \cite{AndreCampos2015}.

The second eigenvalue lays down additional constraints to ensure stability of the solution
\begin{align}
	\lambda_2 < 1  &\Rightarrow \nu < a_C\label{thirdcondition}\\
	\lambda_2 > -1 &\Rightarrow \nu\left(1-\frac{\nu}{a_C}\right) < 2 \Rightarrow \frac{\nu^2}{a_C} - \nu +2 > 0 \label{fourthcondition}
\end{align}
The condition \ref{thirdcondition} has a simple interpretation: if the probability of death exceeds the maximum reproduction rate of the
strain $C$, the solution is no longer stable and the population is doomed to extinction. The last condition (\ref{fourthcondition}) is
always satisfied as $\nu$ is much smaller than one. 

We notice that the stability of the solution with $\hat{n}_D=0$ and $\hat{n}_{C} \neq 0$ is independent of the effect of the 
toxin on the strains, i.e. independent of $\eta$ and $\beta$. This happens because the toxin is only 
produced by strain $D$. Introducing 
a small amount of $D$ will bring an infinitesimal amount of toxin which is not relevant in a first order approximation. 

\begin{figure}[tp]
\centering
\includegraphics[width=0.45\textwidth]{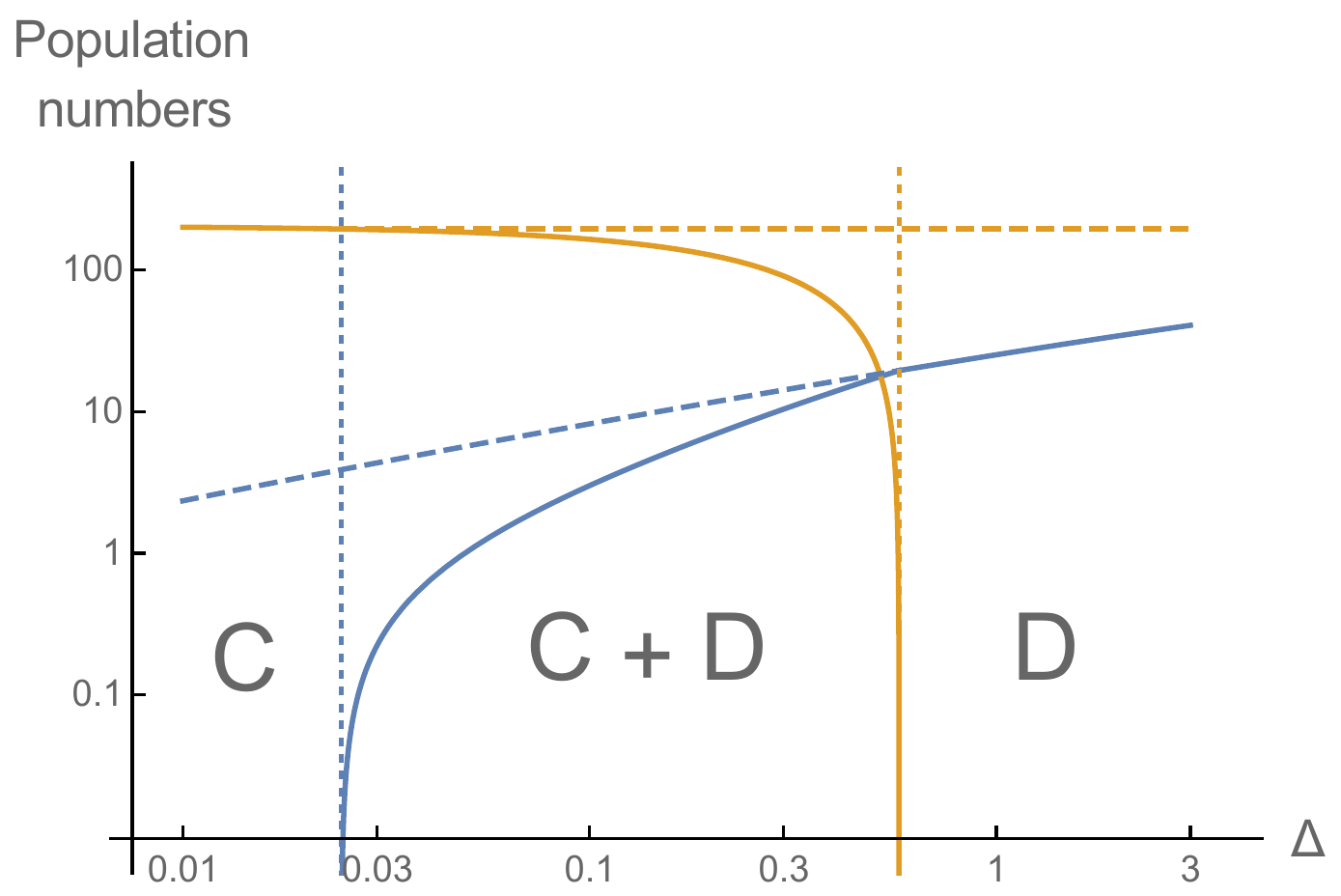}
\caption{Population size as a function of $\Delta$, with $\Gamma=4$, $\epsilon=10$, $S=10$, $a_C=0.2$, $\alpha_C=1$ and $\eta=0.01$. The $C$ population is in yellow and the $D$ population in blue. The full lines denote stable equilibria and the dashed lines unstable equilibria. To the left of blue dotted line a population of pure $C$ is stable and to the right of yellow dotted line a population of pure $D$ is stable. In the middle region the population is stable only in coexistence.}
\label{fig:population_vs_delta_beta0}
\end{figure}

\subsection{Solution III: $\hat{n}_{D} \neq 0$; $\hat{n}_C=0$}
The population size of the inefficient strain $D$ at equilibrium, and in the absence of strain $C$, is given by the solution of
the following equation
\begin{align}
	\exp\left[-\frac{\alpha_D  S}{\hat{n}_D}\right]=1-\frac{\nu +\eta\, \hat{n}_D}{a_D} \label{transcendental-nd}.
\end{align}
The solution of this transcendental equation can be obtained numerically. Since the left-hand side is a monotonically increasing
function of $\hat{n}_{D}$ whereas the right-hand side starts at $1-\frac{\nu}{\alpha_D}$ and monotonically decreases the 
equation has a positive solution whenever $a_{D}>\nu$. When $a_{D}<\nu$  the solution is no longer stable and 
population becomes extinct.

The Jacobian matrix is now given by
\begin{widetext}
\begin{align}
J_{n_C(t)=0}=\left(
    \begin{array}{cc}
     1 -\nu -2 \eta\, \hat{n}_D
     +a_D\left\{1 -\exp\left[-\frac{S\, \alpha_D}{\hat{n}_D}\right] \left[1+\frac{S\, \alpha_D}{\hat{n}_D}\right]\right\} & 
     -\frac{a_D}{\epsilon}\frac{\alpha_D S}{\hat{n}_D} \exp\left[-\frac{S\, \alpha_D }{\hat{n}_D}\right]\\
     0 & 
      1-\nu-\beta \hat{n}_D +a_C\left\{1-\exp\left[-\frac{S\, \alpha_C}{\epsilon\, \hat{n}_D}\right]\right\} \\
    \end{array}
\right)\label{jacobiannc0}
\end{align}
\end{widetext}
where $\hat{n}_D$ is the numerical solution of Eq. (\ref{transcendental-nd}), and its eigenvalues are simply
\begin{align}
	\lambda_1 &= 1 -\nu -2 \eta\, \hat{n}_D
     +a_D\left\{1 -\text{e}^{-\frac{S\, \alpha_D}{\hat{n}_D}} \left[1+\frac{S\, \alpha_D}{\hat{n}_D}\right]\right\}\\
    \lambda_2 &= 1-\nu-\beta \hat{n}_D +a_C\left\{1-\exp\left[-\frac{S\, \alpha_C}{\epsilon\, \hat{n}_D}\right]\right\}.
\end{align}
In Figure \ref{fig:Donly} we numerically solve eq. (\ref{transcendental-nd}) and present in the diagram $\eta$ vs $\beta$ the region
of its stability. As expected, the toxin has a harmful effect on strain $D$ leading to reduced population sizes
 at equilibrium as $\eta$ increases. We observe the existence of a threshold value for $\eta$, $\eta \sim 0.1$, above which the 
population can no longer be sustained. Although the population size of individuals type $D$ is not influenced by the parameter $\beta$ (see eq. \ref{transcendental-nd}), the stability of the solution is severely affected. The larger the effect of the toxin on strain $C$, 
$\beta$, the wider is the region of stability of the solution. The green region, denoting the conditions that ensure the stability
of the present solution is surrounded by two gray areas. The one to the right means that the solution
$\hat{n}_{D} \neq 0$ and $\hat{n}_{C}=0$ is no longer evolutionarily stable, whereas the coexistence solution becomes stable,
as we will see next.  The gray region to the left of the green one is not physically meaningful. 
For those set of values of $\eta$ and $\beta$, the net maximum growth rate of the strain $C$ becomes 
negative, i.e., $1+a_{C}-\nu-\beta\hat{n}_{D}<0$. This occurs because for very small $\eta$ the population size of defectors at equilibrium
can be large, and so does the term $\beta\hat{n}_{D}$.

\begin{figure*}
\centering
\begin{tabular}{ccc}
    \includegraphics[width=0.3\textwidth]{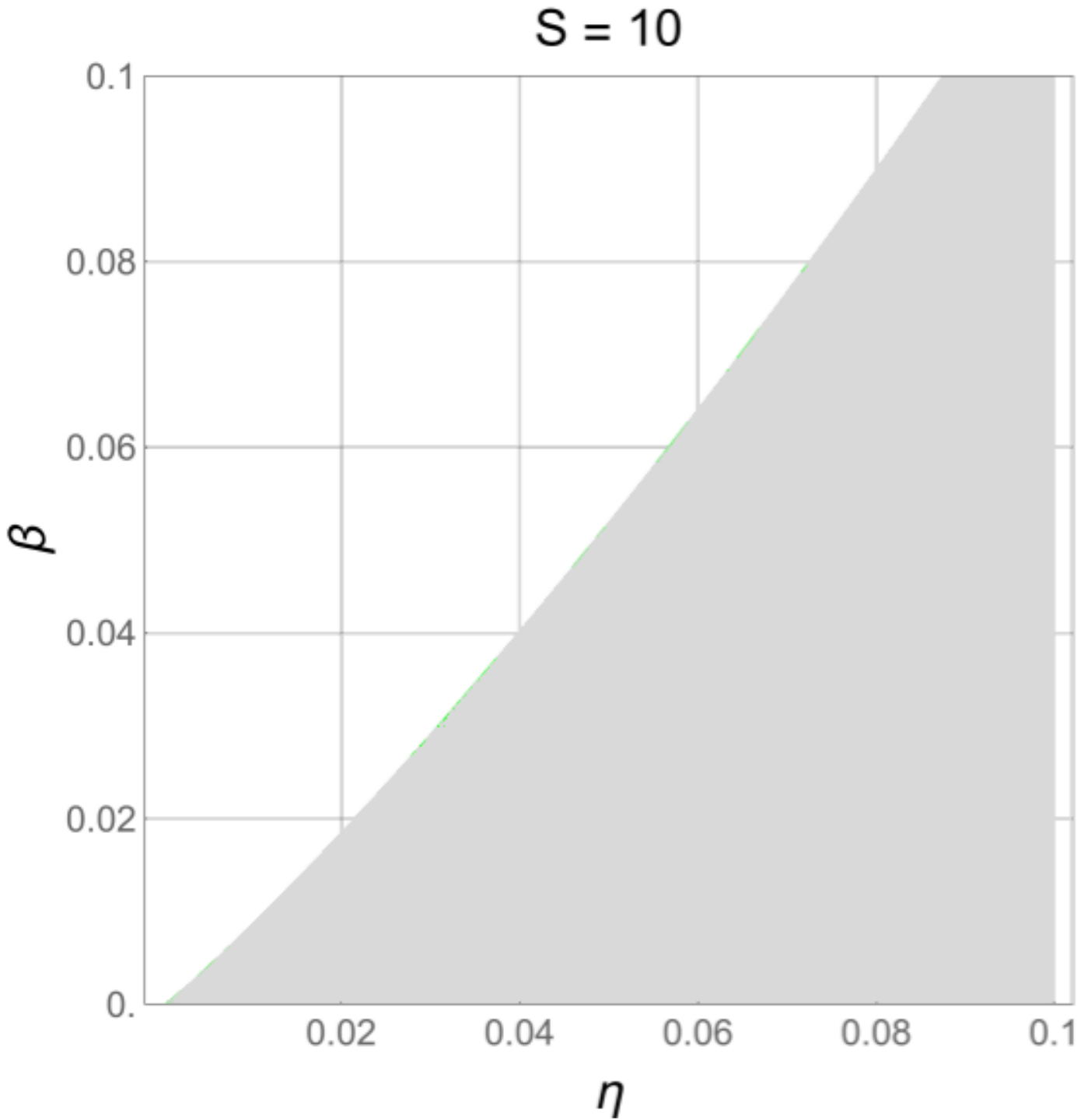}&
    \includegraphics[width=0.3\textwidth]{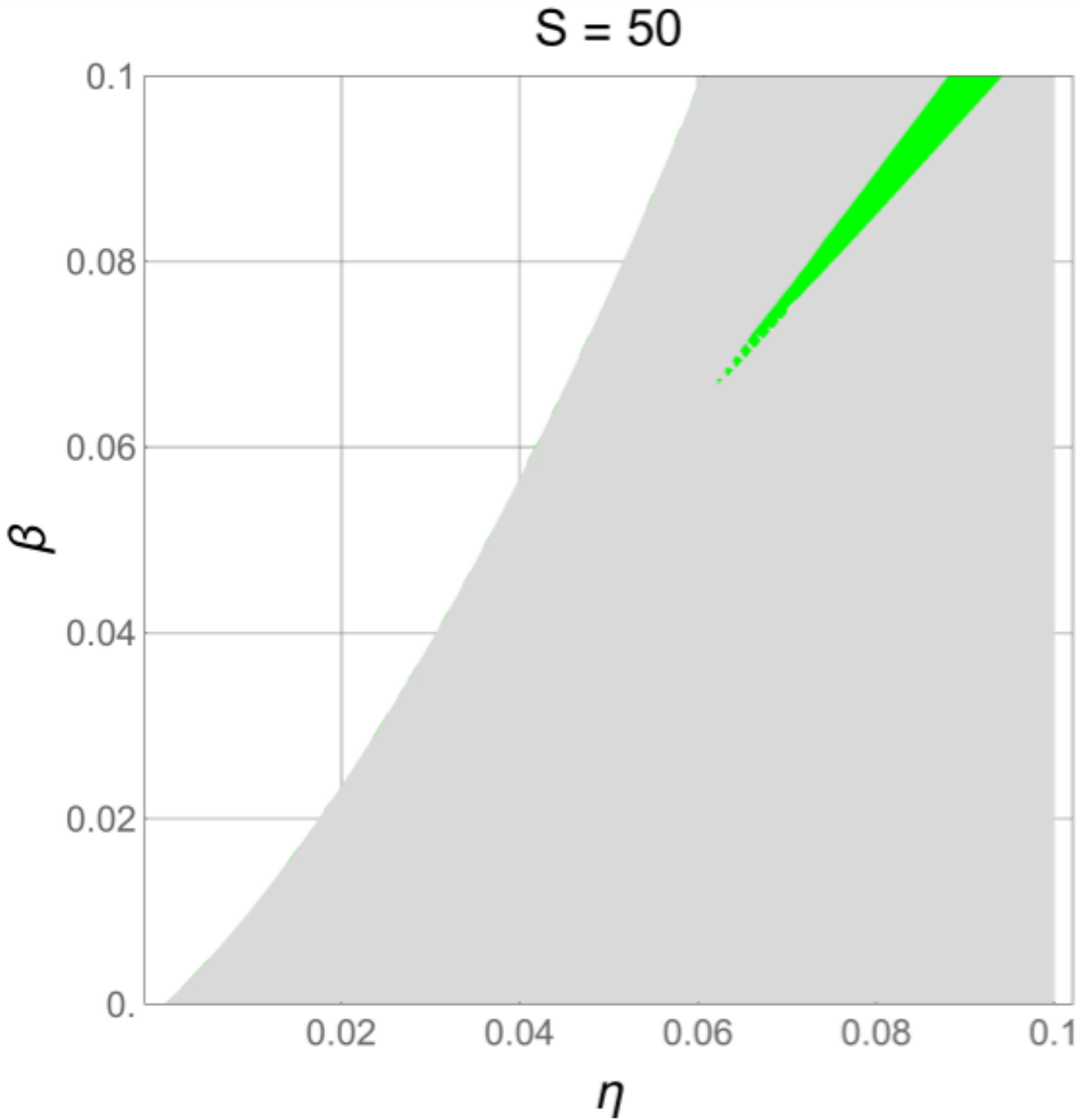}&
    \includegraphics[width=0.3\textwidth]{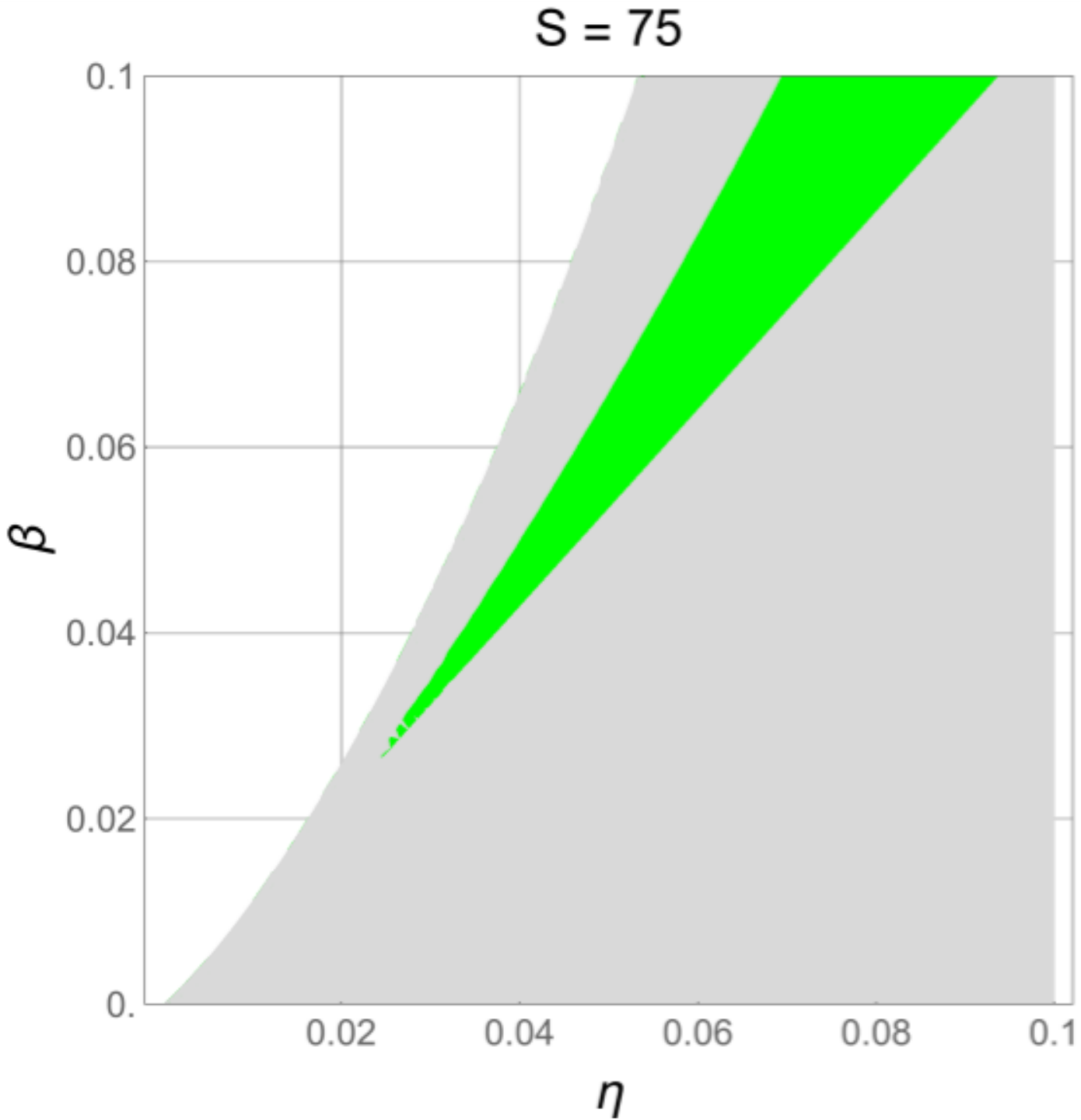}\\
    \includegraphics[width=0.3\textwidth]{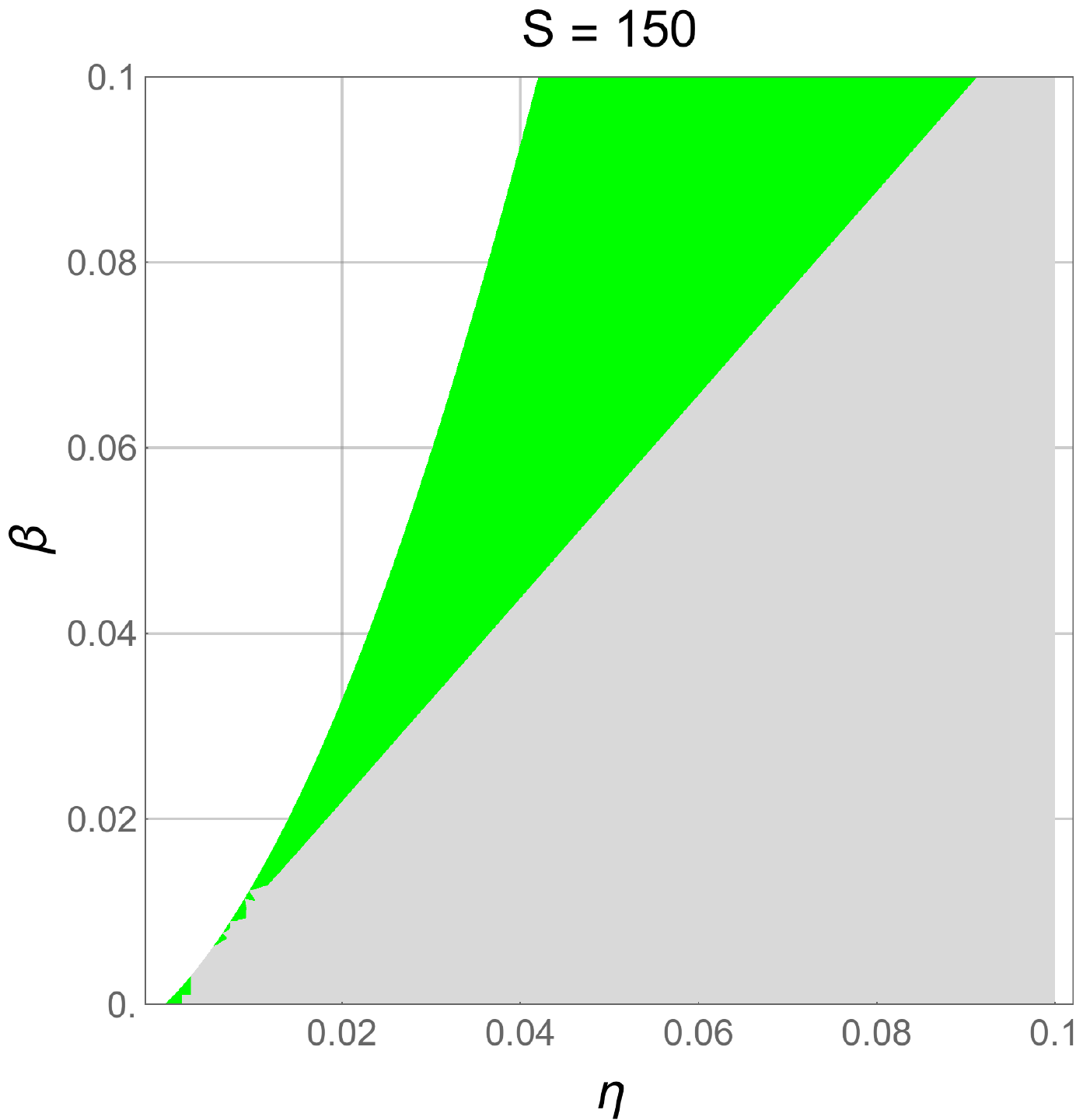}&
    \includegraphics[width=0.3\textwidth]{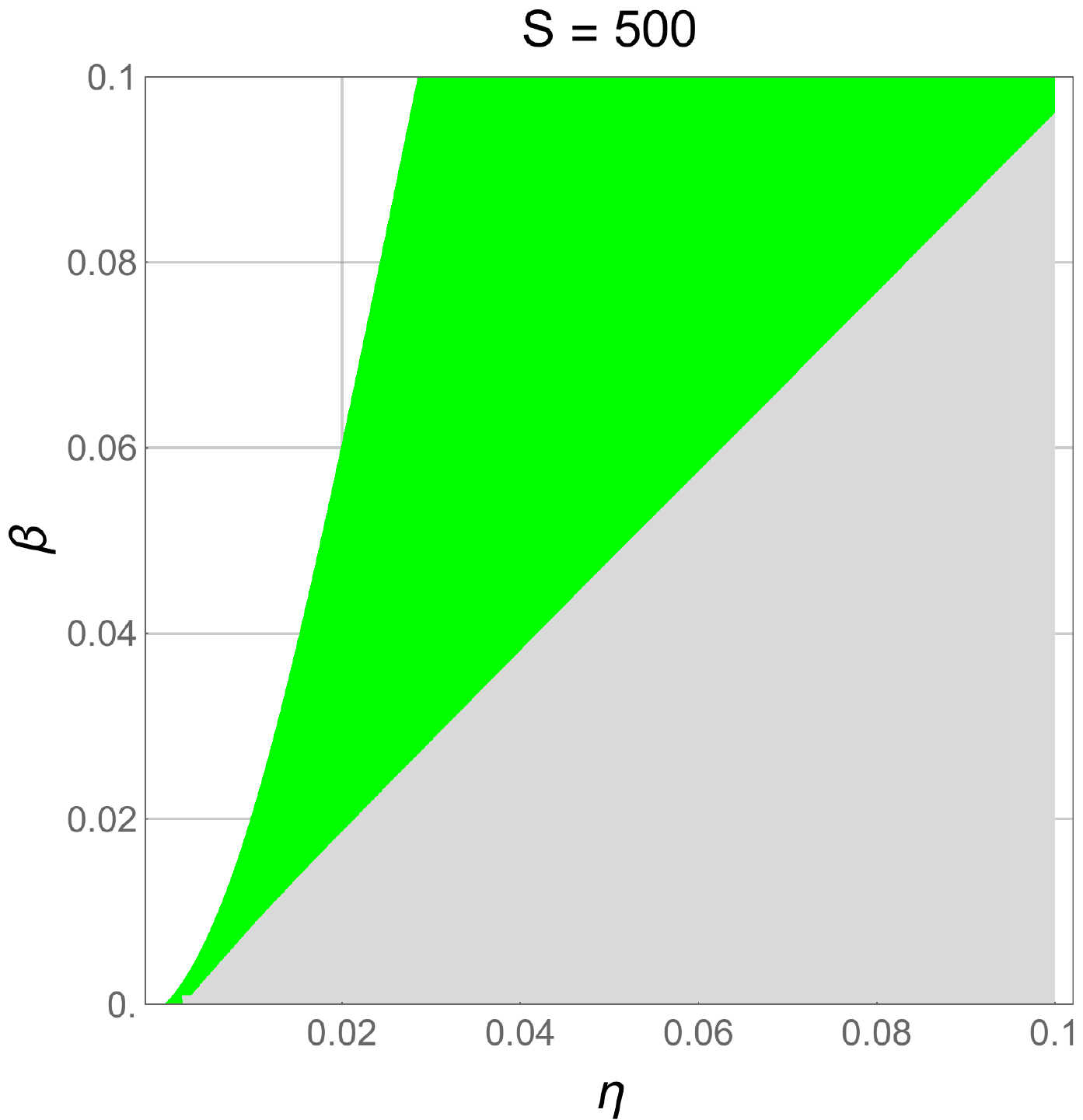}&
    \includegraphics[width=0.3\textwidth]{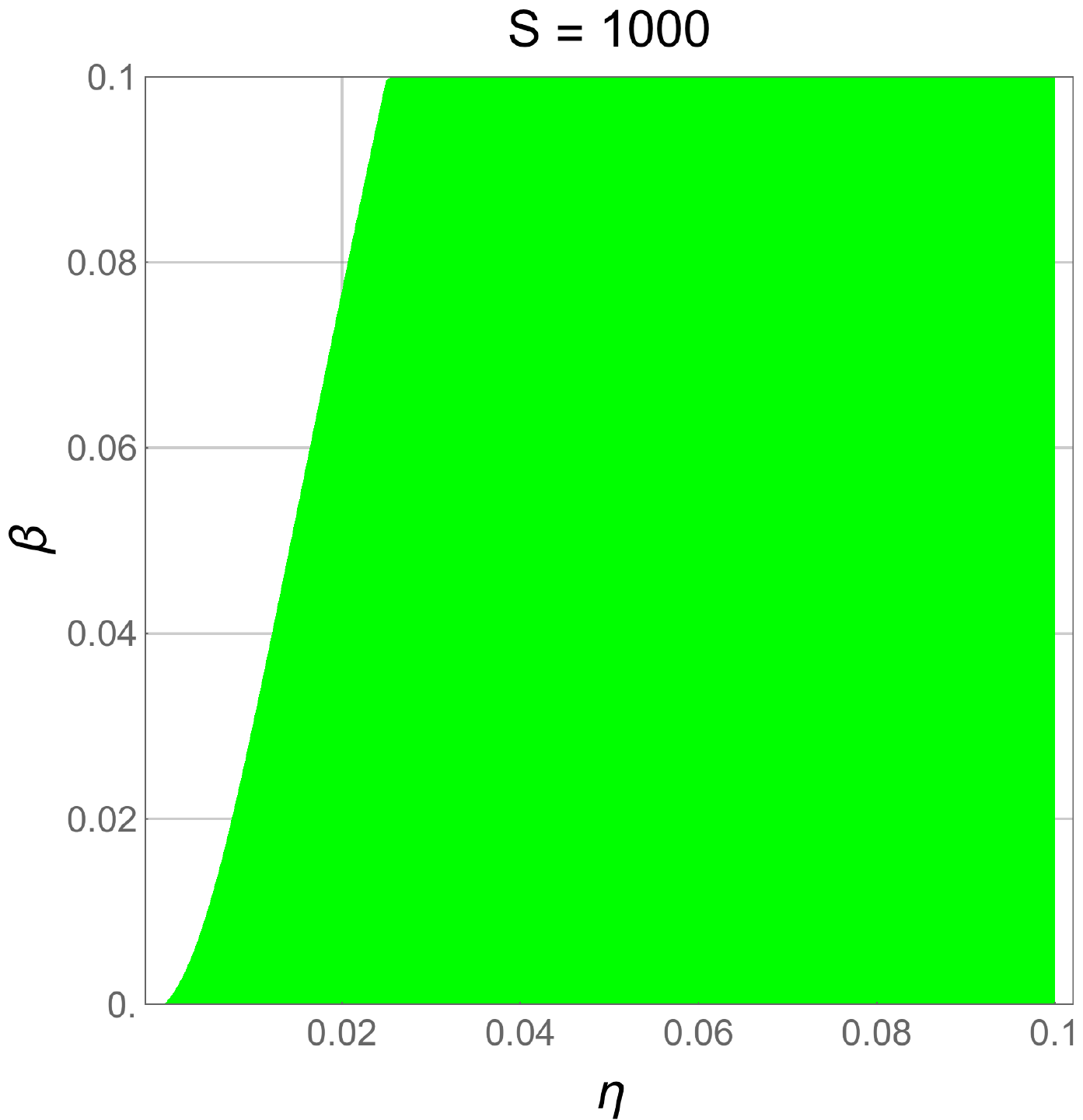}\\
\end{tabular}
    \caption{Region of stability of the coexistence solution ($|\lambda|<1$) for S: 10, 50, 75, 150, 500, 1000 (from left to right, top to bottom). The coloured area denotes the region where the solution is stable, the grey area the region where the solution exists but is not stable. The parameters are $\nu = 0.01$, $\Delta = 0.5$, $\Gamma = 4$, $a_D = 0.2$ and $\alpha_D = 0.5$.}
    \label{fig:stability_vs_sugar}
\end{figure*}

\subsection{Solution IV: coexistence solution}
The system allows another type of solution, where the two strains coexist at equilibrium, i.e., $\hat{n}_{C}\neq 0$ and 
$\hat{n}_{D} \neq 0$. This new equilibrium, which is in practice missing in the lack of the toxin (see Ref. \cite{AndreCampos2015}), 
is found by the solving the following pair of equations
\begin{align}
	\left[1-\frac{\nu +\beta \hat{n}_D}{a_C}\right]^{\Delta \epsilon}=1-\frac{\nu +\eta  \hat{n}_D}{a_D }\label{coexistence1}\\
	\hat{n}_C=-\frac{\alpha_C\, S}{\log \left[1-\frac{\nu +\beta \hat{n}_D}{a_C}\right]}-\epsilon\,\hat{n}_D\label{coexistence2}
\end{align}

\noindent
In general these equations can only be solved numerically, however, for particular situations analytical equations can be derived. 
As special cases we mention: $\beta=0$ and $\Delta\epsilon = 1$. 

\noindent
The stability of the system can be obtained replacing the numerical solution into the general form of 
Jacobian matrix (please see Appendix A).

\subsubsection{Limiting case $\beta=0$}
This limiting case corresponds to the situation in which the toxin is harmful to its producer but not for strain type $C$. 
By making $\beta=0$ the first equation greatly simplifies, and the equilibrium population of strain $D$ 
becomes
\begin{align}
\hat{n}_D=\frac{a_D}{\eta}\left[1-\left(1-\frac{\nu}{a_C}\right)^{\Delta \epsilon}\right]-\frac{\nu}{\eta}.
\end{align}
The population size of strain $C$ directly follows through the substitution of the above equation into eq. (\ref{coexistence2}).

Figure \ref{fig:population_vs_delta_beta0} displays a plot of the population size of both strains versus the ratio 
$\Delta =\frac{\alpha_{D}}{\alpha_{C}}$, which quantifies the relative efficiency of strain $D$ over strain $C$. For small $\Delta$, meaning
that strain $D$ has a poor yield, the population is only comprised by individuals of type $C$ at equilibrium. The onset of the
coexistence takes place at $\Delta \approx 0.025$. At this point the
population of strain $C$ starts to shrink while the population of strain $D$ soars. Although physically meaningful, 
at $\Delta \approx 0.58$, the coexistence solution is no longer stable. Beyond this point the population of strain type $D$ dominates.

\begin{figure*}
\centering
\begin{tabular}{ccc}
    \includegraphics[width=0.3\textwidth]{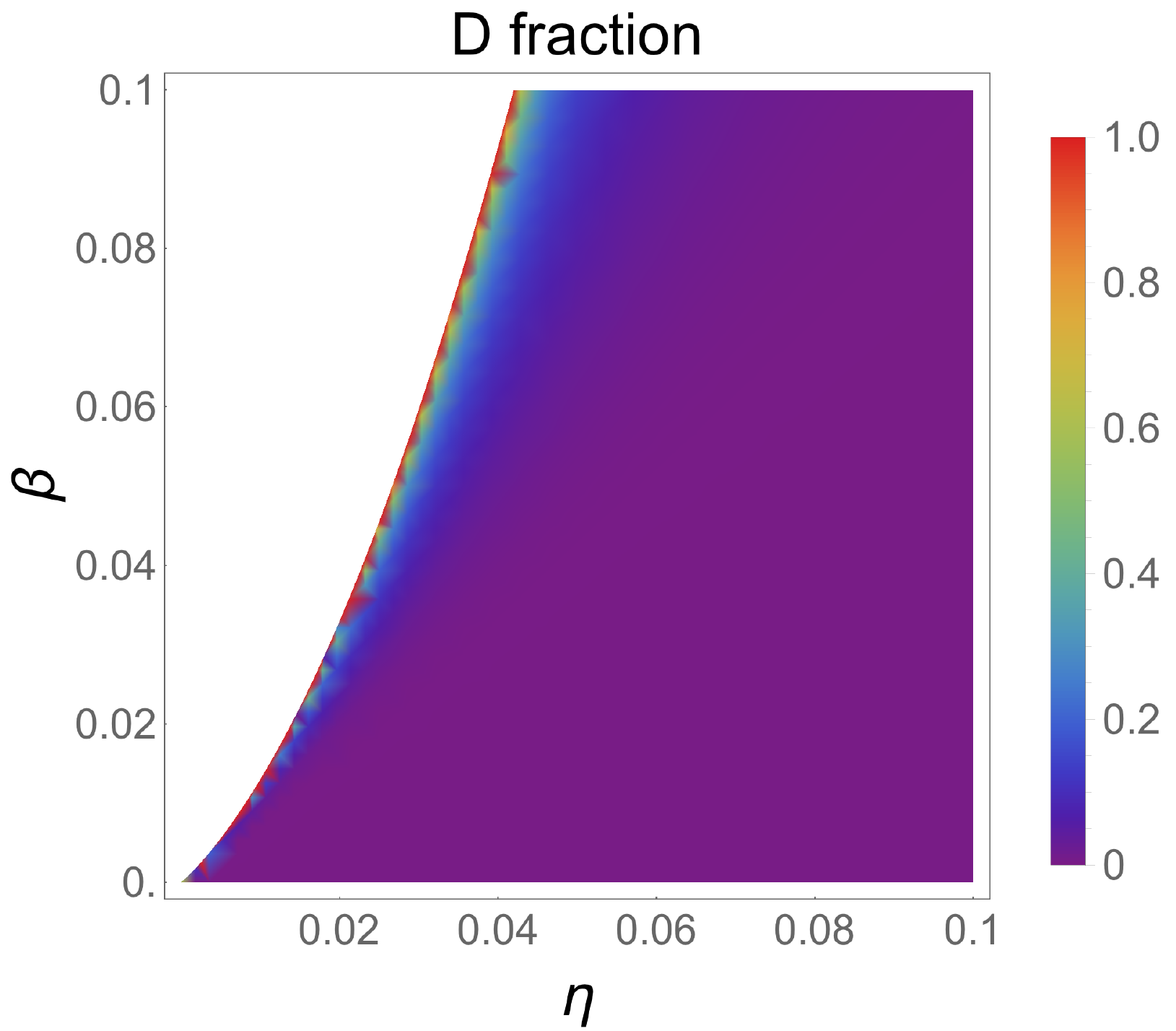}&
    \includegraphics[width=0.3\textwidth]{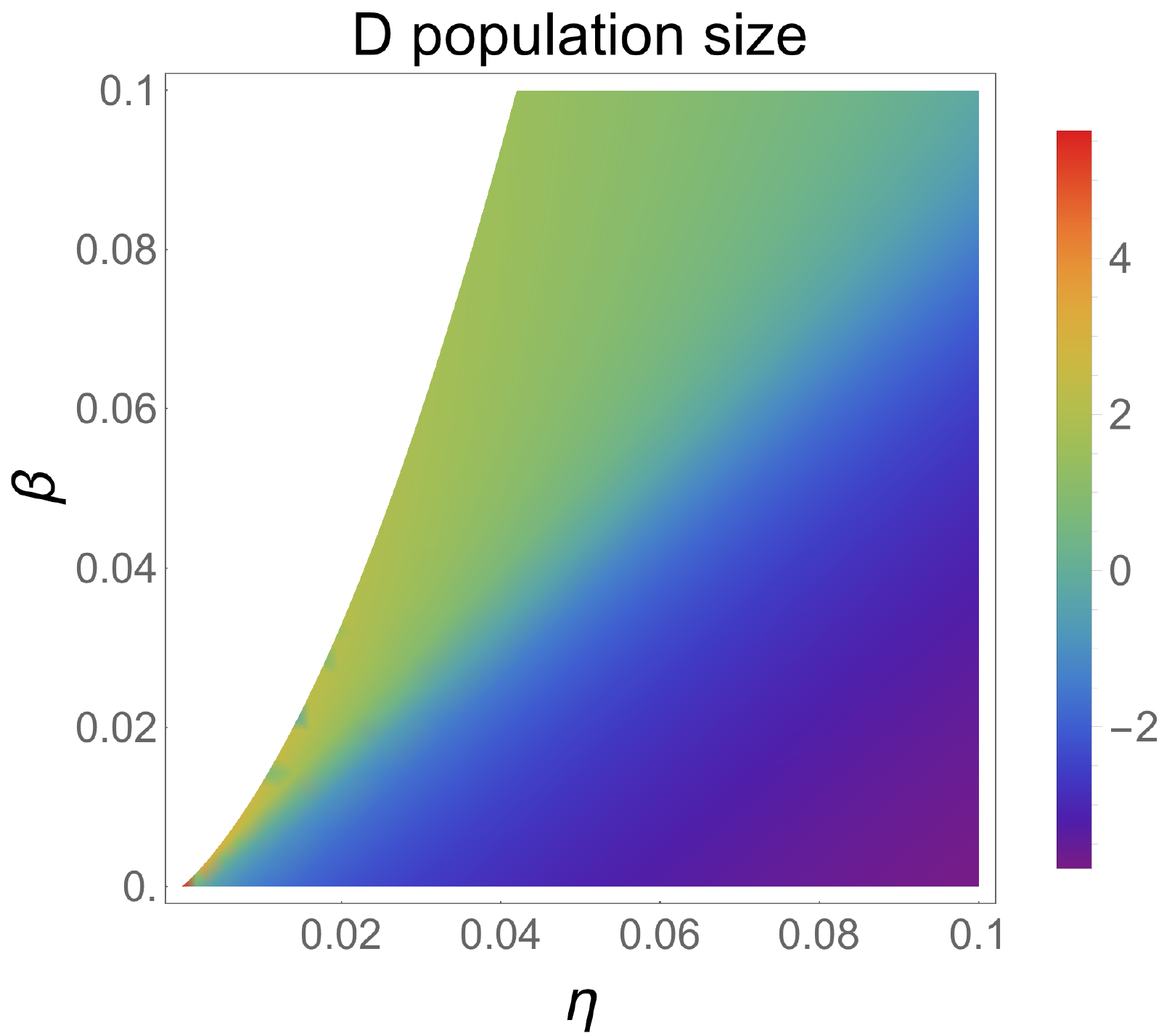}&
    \includegraphics[width=0.3\textwidth]{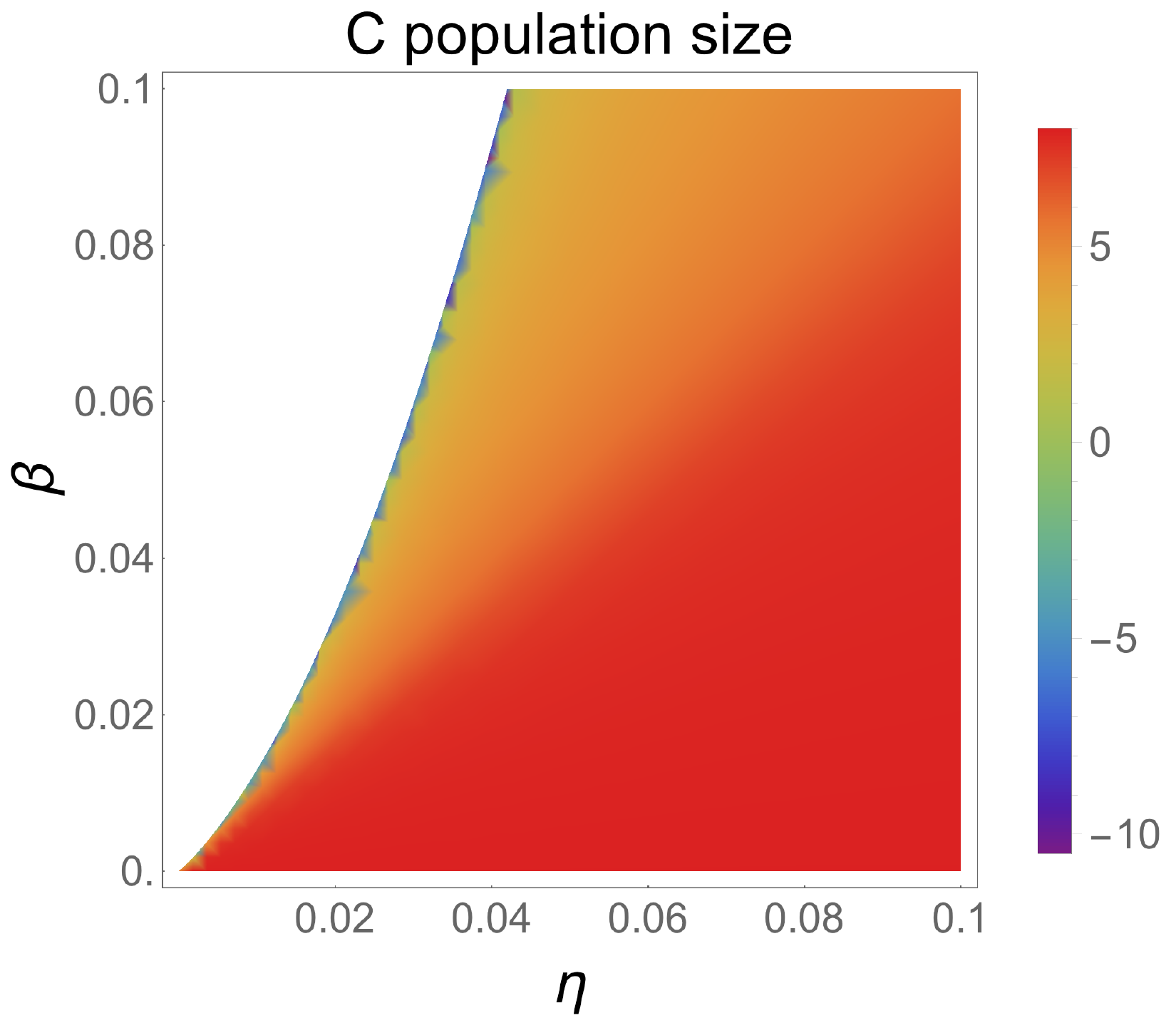}\\
\end{tabular}
    \caption{Left graph: $D$ population fraction. Middle graph: $D$ population size. Right graph: $C$ population size. The parameters are $\nu = 0.01$, $\Delta = 0.5$, $\Gamma = 4$, $a_D = 0.2$, $\alpha_D = 0.5$ and $S=150$. In the middle and right panels the gradation 
is in log scale.}
    \label{fig:population_size_analytical}
\end{figure*}

\subsubsection{The case $\Delta\epsilon = 1$}

There is at least another special case in which an exact coexistence solution can be determined, i.e., $\Delta\epsilon = 1$.
Actually, this is a very restricted situation, because the $\Delta\epsilon = 1$ means that the maximum energy produced by
the efficient strain must be as large as the performance of the strain $D$ by uptaking the resource. This is certainly not the
case for fermento-respiring cells. Though, the condition  $\Delta\epsilon = 1$ enables us to simplify eqs. (\ref{coexistence1}) and 
(\ref{coexistence2}), thus obtaining
\begin{align}
	\hat{n}_D &= \nu\frac{\Gamma-1}{\eta-\Gamma\beta}\\
	\hat{n}_C &=-\frac{\alpha_C\, S}{\log \left[1-\frac{\nu}{a_C}(1 - \frac{1-\Gamma}{\eta/\beta-\Gamma})\right]}-\epsilon\,\nu\frac{\Gamma-1}{\eta-\Gamma\beta}.
\end{align}
It is worth noticing that when $\Delta\epsilon = 1$ and $\eta=\beta$, the coexistence is not possible as in this case 
$\hat{n}_D$ equals 
\begin{align}
	\hat{n}_D &= \frac{\nu}{\eta}\frac{\Gamma-1}{1-\Gamma} = -\frac{\nu}{\eta}
\end{align}
which is always negative and so not physically sound.

\subsubsection{General case}

Now we explore the dependence of the coexistence solution and its stability in terms of the parameters $\eta$ and $\beta$ which
describe the effects of the toxin on the strains. Figure \ref{fig:stability_vs_sugar} shows the regions at which the coexistence 
solution is stable, here represented by the green area. The gray region represents the set of the parameter space at which the
solution still makes sense though it is not stable. In the graph the dependence of the stability region on the amount of
resource is studied. Interestingly, we observe that the stability region grows as the resource influx rate $S$ increases. 
This is a quite striking outcome since in the absence of the toxin, the increase of the amount of resource available to the system
favours the fixation of the selfish strain $D$, while harsh envorinmental conditions tends to favour the cooperative 
trait \cite{PfeifferScience2001,Frick2003,FernandoCamposPRE2013}. This outcome evinces the role played by the toxin production as
the coexistence is even enhanced as the resource becomes more abundant, thus warranting the maintenance of the efficient strain.
In  Figure \ref{fig:population_size_analytical} we see a heat map of the fraction and size of the population of both
strains in terms of $\beta$ and $\eta$ in the set of parameters at which coexistence solution is stable. As expected,
as $\eta$ is increased the fraction and size of the population of the strain $D$ is reduced. The same happening to the population
of strain $C$ as $\beta$ rises. Although we also notice that, for fixed $\eta$, an augment of $\beta$ favours the population of
strain $D$, the opposite situation, fixing $\beta$ and augmenting $\eta$ promotes a much higher variation in the
population size of the efficient strain $C$.

\section*{Conclusions}
We have studied the possible solutions (extinction, only strain C, only strain D and coexistence) of a discrete-time model  
that describes the evolution of two competing metabolic strategies. The model assumes that the population is unstructured.
One of the strains produces a toxin that affects the net growth rate of both strains, yet with different strength.
As empirically found, the model predicts that the coexistence between a high yield strain and a high rate strain is possible. This outcome corroborates the fact that structuring is not the only possible mechanism promoting coexistence.

\bigskip
When the effect of the toxin on the efficient strain is negligible, the coexistence between the two strain types 
is possible for intermediate 
levels of the relative efficiency, $\Delta$, of the inefficient strain. As expected, for very low values of $\Delta$ the 
efficient strain dominates while for high values of $\Delta$ the inefficient strain dominates. The novelty comes about
with the observation of a coexistence regime which takes place at intermediate values of $\Delta$, which 
is lacking if the toxin is not considered \cite{AndreCampos2015}.  

\bigskip

In the following we explored a more general scenario in which the toxin also attenuates the growth rate of the
efficient strain. At this point, the analysis relied on values of relative efficiency $\Delta$ at which the cooperative
strain is not favored in the absence of toxin, and extensively probed the outcome of the model in the parameter space
$\eta$ vs $\beta$. A striking finding of the model is the observation that
the resource influx rate has a prominent role in driving the fate of the population. In a situation where the resource is
scarce the population evolves to a pure population of only individuals of type $D$ (inefficient strain) and the coexistence solution
is always unstable. However, as the 
resource influx rate rises a coexistence domain emerges. The domain of the coexistence solution starts just after the pure strain $D$ 
solution
becomes unstable and widens out as the resource influx rate is augmented. This is quite surprising as the plenty of resource
usually favors the defecting behavior \cite{FernandoCamposPRE2013}. However, although a large amount of resource into the system
means that the inefficient strain ends up seizing a large amount of resource and consequently has an immediate 
 positive effect on the growth rate, subsequently it
also enhances the production of byproducts (toxin). 

Our results corroborates the empirical observation that the coexistence between competing metabolic strategies is a possible outcome,
and proves the effectiveness of the metabolite intermediate as the main driving mechanism for this purpose. Counterintuitively, our
results evince that this coexistence is enhanced in a scenario of plentiful of  resource.

\section*{Acknowledgments}

PRAC is partially supported by Conselho Nacional de Desenvolvimento Cient\'{\i}fico e Tecnol\'ogico (CNPq), and also acknowledges financial support from Funda\c{c}\~ao de Amparo \`a Ci\^encia e Tecnologia
do Estado de Pernambuco (FACEPE). FFF gratefully acknowledges financial support
from FAPESP (Funda\c{c}\~ao de
Amparo \`a Pesquisa do Estado de S\~ao Paulo). JL is supported by Funda\c{c}\~ao de Amparo \`a Ci\^encia e Tecnologia
do Estado de Pernambuco (FACEPE) and AA has a fellowship from Conselho Nacional de Desenvolvimento Cient\'{\i}fico e Tecnol\'ogico (CNPq).






\section*{Appendix A}
The general form of the Jacobian matrix for the analytical approximation is a $2\times2$ with entries
\begin{widetext}
\begin{align}
    J_{11}&=1-\nu + a_D \left\{1-\exp\left[-\alpha_D\frac{\epsilon S}{n_C(t)+\epsilon n_D(t)}\right] \left[1+\alpha_D S\frac{n_D(t) \epsilon^2}{[n_C(t)+\epsilon n_D(t)]^2}\right]\right\}-2 \eta\,n_D(t) \\
    J_{12}&=-n_D(t)\left\{a_D \alpha_D\,S\frac{\epsilon  \exp\left[-\alpha_D\frac{\epsilon S}{n_C(t)+\epsilon n_D(t)}\right]}{[n_C(t)+\epsilon n_D(t)]^2}\right\} \\
    J_{21}&=-n_C(t) \left\{a_C \alpha_C S\frac{ \epsilon \exp\left[-\alpha_C\frac{S}{n_C(t)+\epsilon n_D(t)}\right]}{[n_C(t)+\epsilon n_D(t)]^2}+\beta \right\} \\ 
    J_{22}&=1-\nu + a_C \left\{1-\exp{\left[-\alpha_C\frac{S}{n_C(t)+\epsilon n_D(t)}\right]} \left[1+\alpha_C S\frac{n_C(t)}{[n_C(t)+\epsilon n_D(t)]^2}\right]\right\}-\beta \, n_D(t)
\end{align}
\end{widetext}

\end{document}